# Geomagnetic field influences probabilistic abstract decision-making in humans

Kwon-Seok Chae[1,2,*], In-Taek Oh[2], Soo Hyun Jeong[3], Yong-Hwan Kim[4], Soo-Chan Kim[5], Yongkuk Kim[6]

[1] Department of Biology Education, Kyungpook National University, Daegu 41566, Republic of Korea.
[2] Brain Science and Engineering Institute, Kyungpook National University, Daegu 41566, Republic of Korea.
[3] Department of Baduk Studies, Myongji University, Yongin, 17058, Republic of Korea.
[4] Department of Biological Sciences, Neuroscience program, Delaware State University, Dover DE 19901-2277, United States.
[5] Department of Electrical and Electronic Engineering, Research Center for Applied Human Sciences, Hankyong National University, Anseong 17579, Republic of Korea.
[6] Department of Mathematics, Kyungpook National University, Daegu 41566, Republic of Korea.

* Correspondence: Kwon-Seok Chae (kschae@knu.ac.kr)

## Abstract

To resolve disputes or determine the order of things, people commonly use binary choices such as tossing a coin, even though it is obscure whether the empirical probability equals to the theoretical probability. The geomagnetic field (GMF) is broadly applied as a sensory cue for various movements in many organisms including humans, although our understanding is limited. Here we reveal a GMF-modulated probabilistic abstract decision-making in humans and the underlying mechanism, exploiting the zero-sum binary stone choice of Go game as a proof-of-principle. The large-scale data analyses of professional Go matches and *in situ* stone choice games showed that the empirical probabilities of the stone selections were remarkably different from the theoretical probability. In laboratory experiments, experimental probability in the decision-making was significantly influenced by GMF conditions and specific magnetic resonance frequency. Time series and stepwise systematic analyses pinpointed the intentionally uncontrollable decision-making as a primary modulating target. Notably, the continuum of GMF lines and anisotropic magnetic interplay between players were crucial to influence the magnetic field resonance-mediated abstract decision-making. Our findings provide unique insights into the impact of sensing GMF in decision-makings at tipping points and the quantum mechanical mechanism for manifesting the gap between theoretical and empirical probability in 3-dimensional living space.

## Introduction

To decide the order or selection for taking a turn in games or sports, people often toss a coin as a binary choice (*1–4*), expecting the even probability of the choices for both sides (*1, 2*). The outcome of probabilistic binary choices is implemented in many facets of two-person zero-sum games including poker, chess, and Go, wherein players' choices are frequently mutually exclusive and directly linked to competitive rewards (*5, 6*). In this study, we assessed the stone selection process in Go games as an example of binary choice. Go (called "*baduk*" in Korean) is one of the oldest board games that is mainly popular in East Asian countries (*7, 8*). As an abstract intellectual strategy game, Go has been the most challenging board game for artificial intelligence to defeat professional human players until the recent feat by the computer program AlphaGo (*9, 10*). Since a Go match starts with the first move by a black stone, a stone selection should be conducted before a Go match begins, which is a typical process for deciding who takes the black stone (*11*). This process is important for Go players as it affects the actual winning probability of individual players with black or white stones and both players' game strategies to win (*11–13*). The stone selection is believed to be a fair binary choice because both players participate in either of the two decision steps, and the probability for each stone is regarded as 0.5. However, it is unclear whether the empirical probability is equal to the theoretical probability in binary choices as people easily assume. For instance, the scientific understanding of coin toss probability is limited, and earlier studies stand in a sharp contrast to the belief as the deterministic to be unequal (*14–16*) or a quantum phenomenon with theoretically even odds (*17, 18*).

There is a mount of supporting evidence that the geomagnetic field (GMF) plays an important role as a sensory cue for long- or short-distance migration (*19–24*), body alignment (*25–27*), food foraging (*23, 24*), and magnetic imprinting (*24*) in numerous species. These GMF-sensitive movements rely on directional magnetic information including inclination (*28, 29*) or polarity compass (*30, 31*) derived from GMF nearby. Studies on human magnetoreception of the GMF are relatively rare, and the results that humans can sense GMF have been controversial (*32–35*). Recently, two studies demonstrated that human males can sense GMF by a blue light (*36, 37*) via magnetic field resonance-dependent mechanism (*37*) and then orient the movements toward four cardinal (*36*)- or two alternative (*37*)- magnetic directions through a non-canonical inclination compass (*37*). Indeed, humans have a spin chemistry-based radical pair mechanism (*38–40*)-involved putative magnetoreceptor cryptochrome protein in the eyes (*41*), similarly as GMF-sensitive migratory birds (*42, 43*). However, the impact of GMF other than the needs for physical movements in animals or humans has barely been reported, despite some cognitive implications have been suggested (*44, 45*).

Animals constantly face environmental changes to adopt the most favorable option for enhancing the odds of their survival (*46*). Likewise, humans can make near-optimal decisions in unfavorable two alternative forced-choice tasks using largely unknown strategies (*47, 48*). We postulated that humans might have evolved to incorporate alternative directional information from GMF sensing (*37*) into an alternative abstract decision-making framework, when ordinary sensory modalities such as the five senses are not reliable. This magnetic framework could be helpful for making optimal decisions in probabilistic choices. To test this hypothesis, we adopted the zero-sum stone choice of Go games as an empirical and experimental paradigm to investigate the potential implication of GMF in binary decision-making as a cause of the potential discrepancy between theoretical and empirical probability in binary choice and the underlying mechanism therein.

## Results

### The discrepancy between the empirical and theoretical probability of binary stone choices in Go games

In order to examine whether empirical probability equals to theoretical probability, we initially collected large-scale stone selection data from professional matches between January 2000 and February 2018 supervised by the Korea Baduk Association (*49*), an organization that oversees professional Go games in South Korea. About the rules for a stone selection, procedural descriptions are illustrated visually in Fig. 1, A–C and Movie S1 (details in Materials and Methods). It is assumed that both players have the equal theoretical chance of selecting black or white stones because they are engaged in the decision process without knowing their stone colors, until the number of white stones is revealed to be odd or even. We analyzed 21,212 Go final-round matches held during the aforementioned period (Table S1), with 311 players listed in the player ranking as of June 2018 (see Materials and Methods). Notably, final-round matches are more important and competitive than preliminary matches for professionals, because of the prize money and enhancement of ranking. As expected, the winning rate was associated with higher ranking for all players (Fig. S1A and Data S1). However, contrary to the expectation that the black (and white) stone rate would be about 50%, irrespective of player ranking, there was a significant positive correlation between the black stone rate and higher player ranking in the entire data set of the period aforementioned (Fig. 1D and Data S2) and in a set of year-based data between 2009 and 2017 (Fig. S1B and Data S1). Note that the black stone winning rate of a player is considered to be equivalent to the "success" rate of a player according to the stone selection rule. These results provide compelling evidence that the rate of black stone selection in professional Go matches was not equal and significantly higher for upper-ranked players than lower-ranked players. To ascertain the unequal empirical probability in the stone selection, we conducted stone choice games *in situ* by the same rule above between amateur Go players at special Go schools or Go training academies where they attended for years to become professional Go players (see Materials and Methods). Our results also support that players with upper master levels had significantly higher actual chances for selecting black stones than players with lower master levels (Fig. 1E and Data S2). These results suggest that the empirical probability of binary stone choices in Go games may be remarkably unequal to the theoretical probability of 0.5 (Fig. 1, C–E).

**Geomagnetic field influences the empirical probability of probabilistic abstract decision-making**

Due to the intriguing results above, we investigated the underlying mechanism for the discrepancy between theoretical and empirical probabilities by exploiting the stone choice paradigm as a proof-of-principle. In order to test this paradigm in the setting of modulated GMF, a player's head was situated to locate around the core of the Helmholtz coils (north seat) in a three-dimensional space, and the other player sat on a chair outside of the coils (south seat) of the magnetic north-south axis (Fig. 2A, see Materials and Methods). Note that the ambient GMF at the north seat but not at the south seat was markedly modulated by the Helmholtz coils, depending on the experimental conditions (Table S2). The players were allowed to see the score board freely throughout the experiments (Fig. 2A and Fig. S2A). All the data analyses were based on the black stone winning rate (%) for north seat players in the zero-sum game paradigm. Therefore, both players were provided with even experimental conditions by randomization and taking turns for fair games. In addition, based on our previous findings that short-term fasting is a prerequisite for men's geomagnetic sensations (*36, 37*), we assessed the fasting effects on potentially differential gender responses with fasting male and female subjects for up to 20 h. Then, we performed the stone selection (20 trials between the same players at each seat/game) with a transparent screen set first between the players (see Fig. S2B and Movie S2). The black stone winning rate (the index of "success" in the stone choice) for the north seat players was appreciably decreased for men by the cancellation of GMF to near-zero intensity (near-zero GMF), compared to that under the ambient GMF, whereas it was not different for women (Fig. 2B and Data S3). In contrast, the near-zero GMF significantly reduced the rate for men but not for women with an *opaque* screen combined with ear-muffing (Fig. S2C) to eliminate

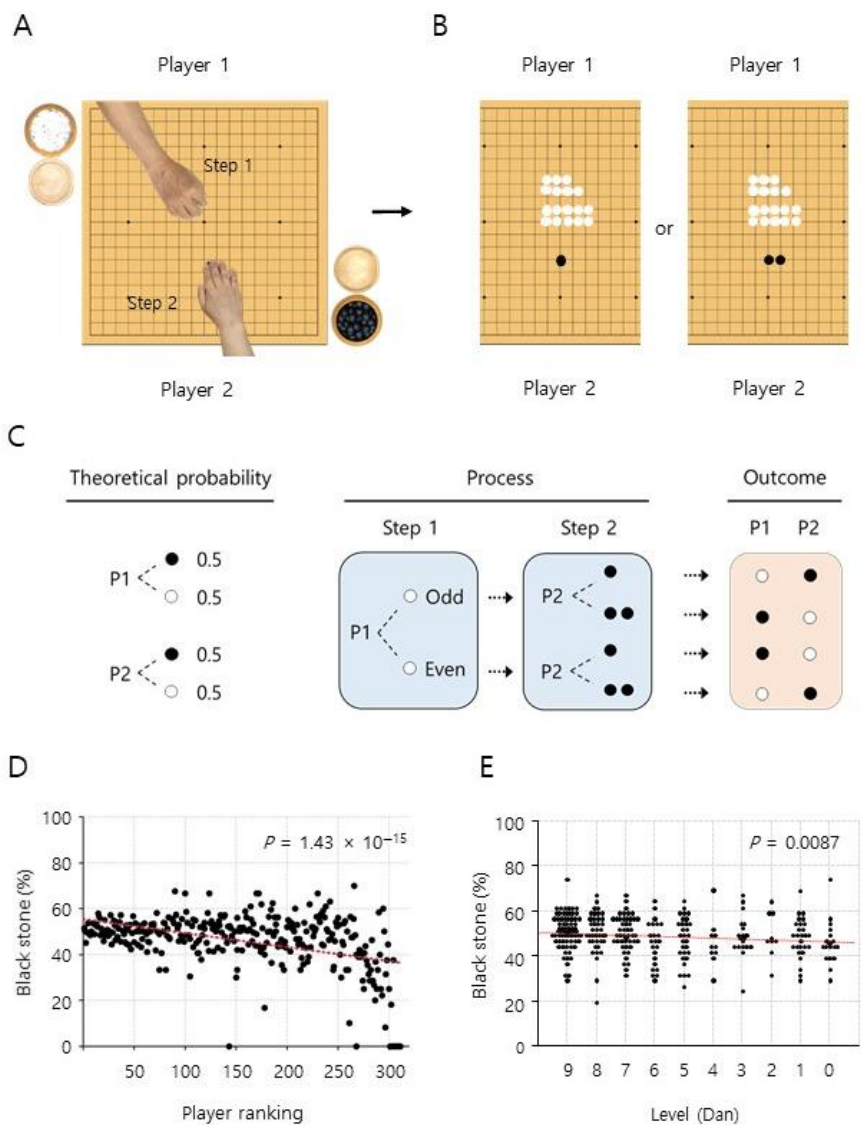

**Fig. 1. Discrepancy between the empirical and theoretical probability of binary stone choices in Go games.**

(**A**) In the step 1, the player 1 grabbed a handful of white stones and placed them on the board while keeping the stones covered. In the step 2, the player 2 placed one or two black stone(s) on the board. (**B**) The player 1 revealed whether the number of white stones (17 in this example) was odd or even. (**C**) The theoretical probability for both players to choose either the black or the white stone is equal. In the flow-chart for the process and outcome of stone choice, dashed lines, or arrows indicate the possible routes for the probable cases in stone choice. If the number of the white and black stone(s) on the board is matched to be either odd or even, players keep the same color of the stones adjacent to them. If not, they switched stone colors for the match. P1, player 1; P2, player 2; white circles, white stones; black circles, black stones. (**D**) A linear fit analysis on the data from the final rounds of 21,212 Go matches among 311 players over 18 years is shown to be inversely correlated between players ranking and black stone rate. $y = -0.061 x + 55.51$, $P = 1.43 \times 10^{-15}$ (except outliers ($n = 11$, 3.5%) with a 0% black stone rate due to having no black stone selection in any final-round matches, $P = 9.38 \times 10^{-10}$), Pearson's $r = -0.43$, $n = 311$. The black stone rate of higher-ranked players (1–150th) (49.6 ± 7.1%, mean ± SD) vs. lower-ranked players (151–311th) (42.7 ± 15.6%, mean ± SD), $P = 3.5 \times 10^{-7}$ was analyzed by a two-sample *t*-test. (**E**) The analysis of *in situ* stone choice games among 293 amateur players shows a significant correlation between master level (Dan) and black stone rate by the linear fit analysis. $y = 0.483 x + 47.10$, $P = 0.0087$, Pearson's r = 0.15, $n = 293$.

any potential visual or auditory effects by the opponent or surroundings (Fig. 2C and Data S3). The near-zero GMF additionally induced a significant decrease in the rate for women but not for men under the same opaque screen condition (Fig. 2D and Data S3) after a short fast (~ 7 h), in which fasting duration was not tested in our previous studies (*36, 37*). Interestingly, the same glucose level in the blood (mean value, 5.1 mM) under different fasting conditions for men (~ 20 h, Fig. 2E and Data S3) and women (~ 7 h, Fig. 2F and Data S3) suggests a potential causal relationship between the glucose level and the reduced black stone winning rates under the near-zero GMF (Fig. 2, C, and D). Taken together, empirical probabilities in the binary choices were significantly affected by the GMF and food context for men and women differently, revealing that GMF plays a role in affecting the probabilistic outcomes of binary decision-making in humans.

**Geomagnetic sensing may affect intentionally uncontrollable decision-making**

To identify the influencing target of GMF in the binary decision-making, we conducted a systematic analyses of data from male subjects with ~ 20 h fast under the opaque screen condition (Fig. 2C). Under these conditions, potential influences including visual or auditory cues from the opponent and/or surroundings were eliminated, and the effect of near-zero GMF was a remarkable factor for black stone winning rate. First, a time series analysis in the black stone winning rate of the north seat players by trial showed that the near-zero GMF produced a tendency of reduction in the rate up to the 12th trial compared to controls, despite a lag in the first two trials, while the trend from the 13th trial was random (Fig. 3A and Data S4). A stepwise analysis showed that the rate tends to decrease by the near-zero GMF in step 1 and 2 (−10.3% for each step compared to the corresponding control), supporting that GMF can be a contributing factor for the decision-making in both steps (Fig. 3B and Data S4). After considering the process of stone selection in Fig. 1C, we attempted to understand the mechanistic principle how GMF affected the step 1 and 2. Even though it is not statistically significant, the odd/even rate of white stones under the control condition showed a tendency of being reversed by the near-zero GMF in the step 1 (Fig. 3C and Data S4), whereas the odd/even rate of black stones in the step 2 was not changed by the same treatment (Fig. 3D and Data S4). Notably, the subjects answered in the post-experiment questionnaire that they were not able to intentionally grab white stones to be odd or even in the step 1 (approximately 99%), whereas they could intentionally choose one or two black stone(s) in the step 2. Strikingly, the black stone rate in the odd cases under the control was significantly diminished by the near-zero GMF in the step 1 but not in the step 2 (Fig. 3, E and F, and Data S4). In contrast, the changes by the near-zero GMF in the even cases of the step 1 and 2 were not significantly detected (Fig. 3, E and F, and Data S4), suggesting that the GMF was mostly influential in the odd cases of the step 1 to produce the discrepancy between the empirical and theoretical probabilities. These results support that GMF more likely affects intentionally uncontrollable probabilistic binary decision-making (step 1 in this process) than intentionally controllable decisions (step 2).

**The continuum of GMF lines between players is crucial to the novel magnetic field resonance-mediated abstract decision-making**

To investigate the underlying mechanism of GMF-

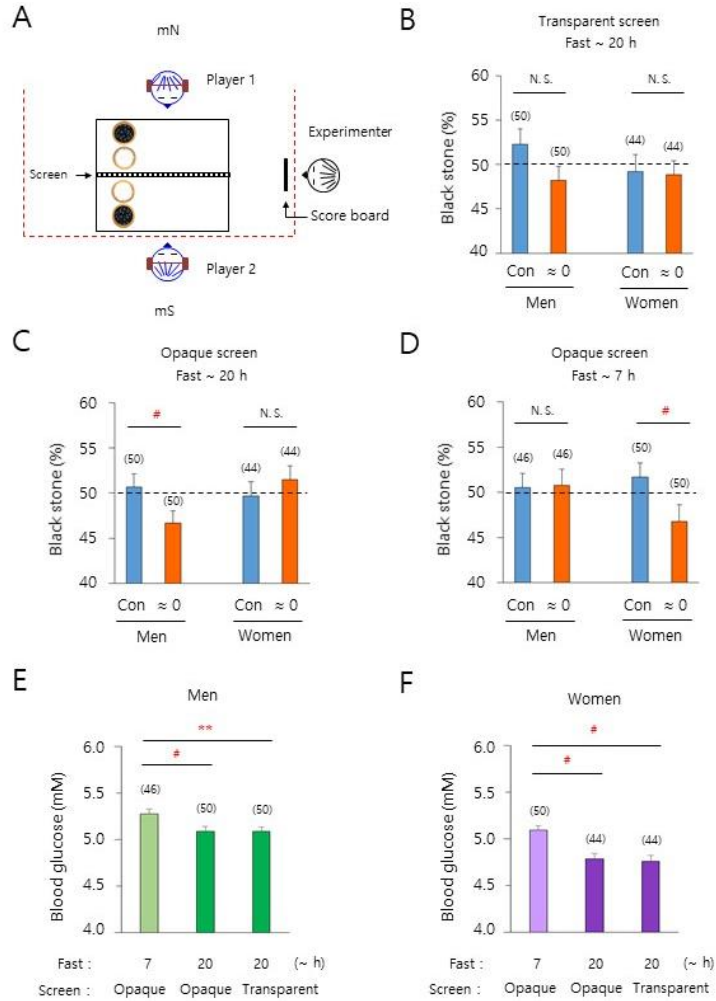

**Fig. 2. Geomagnetic field influences the empirical probability of probabilistic abstract decision-making.**

(**A**) A schematic drawing of the experimental setup for the stone choice paradigm. A top view of two players facing each other across the board with an experimenter who sat east side from the players to conduct experiments. Two players swapped the seat before the second set of a game by the experimental procedure. A transparent or opaque screen stood on the midline of the board between the players; the latter was combined with the ear-muffing of the players. A scoreboard was maintained by an experimenter, so that real-time score was noticed by the players. mN and mS, the ambient magnetic north and south direction, respectively; rectangular solid line, the contour of the board; red dashed line, the outline of the symmetric lower half of Helmholtz coils for the vertical axis and the upper half was omitted for simplicity; black and white circles, two sets of stone buckets on the board for each player at both seats. (**B** to **D**) Black stone winning rates under different experimental conditions that are indicated above each of the graphs. Note the significant differences in the black stone rate in men (C) and women (D). (**E** and **F**) Blood glucose levels in different fast conditions were determined shortly before the first game of a stone choice experiment. Con, control (the ambient GMF); ≈ 0, near-zero GMF; statistical values, mean ± standard error of the mean (SEM); N.S., not significant and $^{\#}$, $P < 0.025$, by the percentile bootstrap analysis; **, $P < 0.01$ by a two-sample $t$-test; horizontal dashed lines, the theoretical probability (50%) for black stone rate. The number of subjects for each group is indicated above bar graphs.

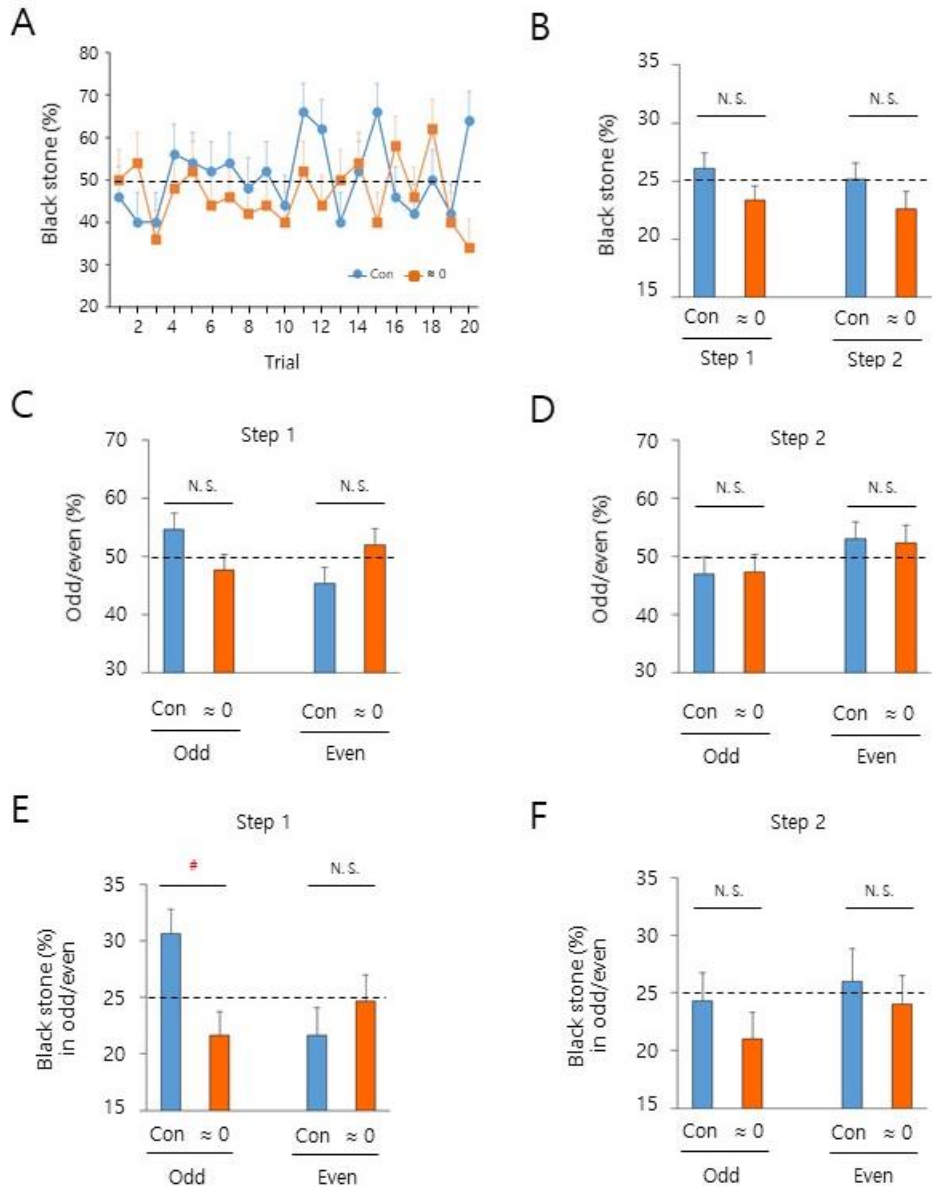

**Fig. 3. Intentionally uncontrollable decision-making was preferably affected by geomagnetic field.**

(**A**) A profile of the black stone rate for the north seat players was displayed by a time series analysis. The same data in Fig. 2C from men were analyzed by trial. Note a continued reduction of the rate by the near-zero GMF, compared to the control, up to the 12th trial with a lag at the first two trials. (**B** to **F**) The different rates of the north seat players by a stepwise analysis. The data of 1–12th trials were analyzed for the black stone rate, odd/even rate, or black stone rate in odd/even by step (see Materials and Methods). Note the significant reduction of the black stone rate by the near-zero GMF in the odd cases of the step 1 (E). Con, control (the ambient GMF); ≈ 0, near-zero GMF; N.S., not significant and #, $P < 0.025$, by the percentile bootstrap analysis; horizontal dashed lines, 25% or 50% theoretical probability for each y-axis index of the graphs; error bars, SEM. Subjects for each Con or ≈ 0 group in the panels (A) to (F) are based on $n = 50$.

modulated decision-making, the same experimental conditions were employed as applied for Fig. 3. Given the magnetic field resonance-dependent magnetoreception in humans (*37*), the electron Larmor frequency (*40*) (1.260 MHz, radiofrequency 1: RF 1) disrupting GMF sensing, was provided vertically (37° relative to the ambient GMF) to the north seat players (*37*). The black stone rate was significantly reduced by the frequency (RF 1), but not by the non-resonant frequency (1.890 MHz, RF 2) (*37, 40*), indicating that a magnetic field resonance mechanism mediated the GMF-modulated decision-making (Fig. 4A and Data S5). The same time series and stepwise analyses performed in Fig. 3 revealed that the RF1 specifically disrupted the GMF-modulated decision-making in a very similar way as the near-zero GMF (Fig. S3 and Data S6), confirming that GMF sensing notably affected the intentionally uncontrollable decision-making step 1, and more effectively influenced the odd cases of the step. As an unbiased approach, we dissected the role of each component of the GMF in decision-making. First, the magnetic field comprised of horizontal component only that oriented toward the magnetic north with the same intensity as the ambient GMF (horizontal only mN, Fig. 4, B, and C) (*28, 37*) did not show a notable difference compared to the control (Fig. 4D and Data S5). Additionally, the inversion of the vertical component of GMF, a diagnostic tool for the involvement of an inclination compass (*20, 37*), also showed no difference compared to the control (Fig. 4D and Data S5). These results strongly suggest that the conventional radical pair mechanism that requires the vertical component as a directional reference (*20*) or an inclination compass, is not necessary for the abstract decision-making.

Likewise, the horizontal component only oriented toward the magnetic south (mS), east (mE) or west (mW) was tested. The mS (Fig. 4, B, and C) was not significantly different from the mN control, implying that a polarity compass in the N-S axis (*30, 31*) may not be involved in the mechanism (Fig. 4E and Data S5). Intriguingly, both mE and mW (Fig. 4, F, and G) where the magnetic field lines passed through north seat players on the magnetic east-west axis and thus perpendicular to those of south seat players, produced a significant decrease in the rate compared to the mN control (Fig. 4E and Data S5). In contrast, the horizontal component only directed toward the magnetic northeast (mNE) or northwest (mNW) that was 30° east or west of the magnetic north, respectively, did not show notable differences, compared to the same mN control (Fig. S4 and Data S7). For better understanding of the results from the "horizontal only" magnetic fields above, note the directions of dashed arrows or arrowheads around north seat player (P1)'s head. The black stone winning rate for P1 was significantly decreased compared to the corresponding control, only if the modulated magnetic field lines around P1's head were perpendicular to the ambient GMF lines as in Fig. 4, F, and G. Nonetheless, the ambient GMF and each of the modulated magnetic fields lines surrounding the P1 and P2 (south seat player) eventually connect as a whole, regardless of the differences in the modulated magnetic fields at the location of P1's head. Similarly, we tested the vertical component with the same intensity as the ambient GMF. Either the downward or upward vertical component only (Fig. 4H) was unexpectedly sufficient to produce a comparable rate compared with the ambient GMF control (Fig. 4I and Data S5), suggesting that magnetic connectivity between players through the vertical components-relayed global circuits of the GMF lines enabled the GMF-modulated decision-making. These results support that the continuum of magnetic field lines between players, but not the intensity or polarity of the magnetic field, was critical for the *outward* influence of players in GMF-modulated decision-making.

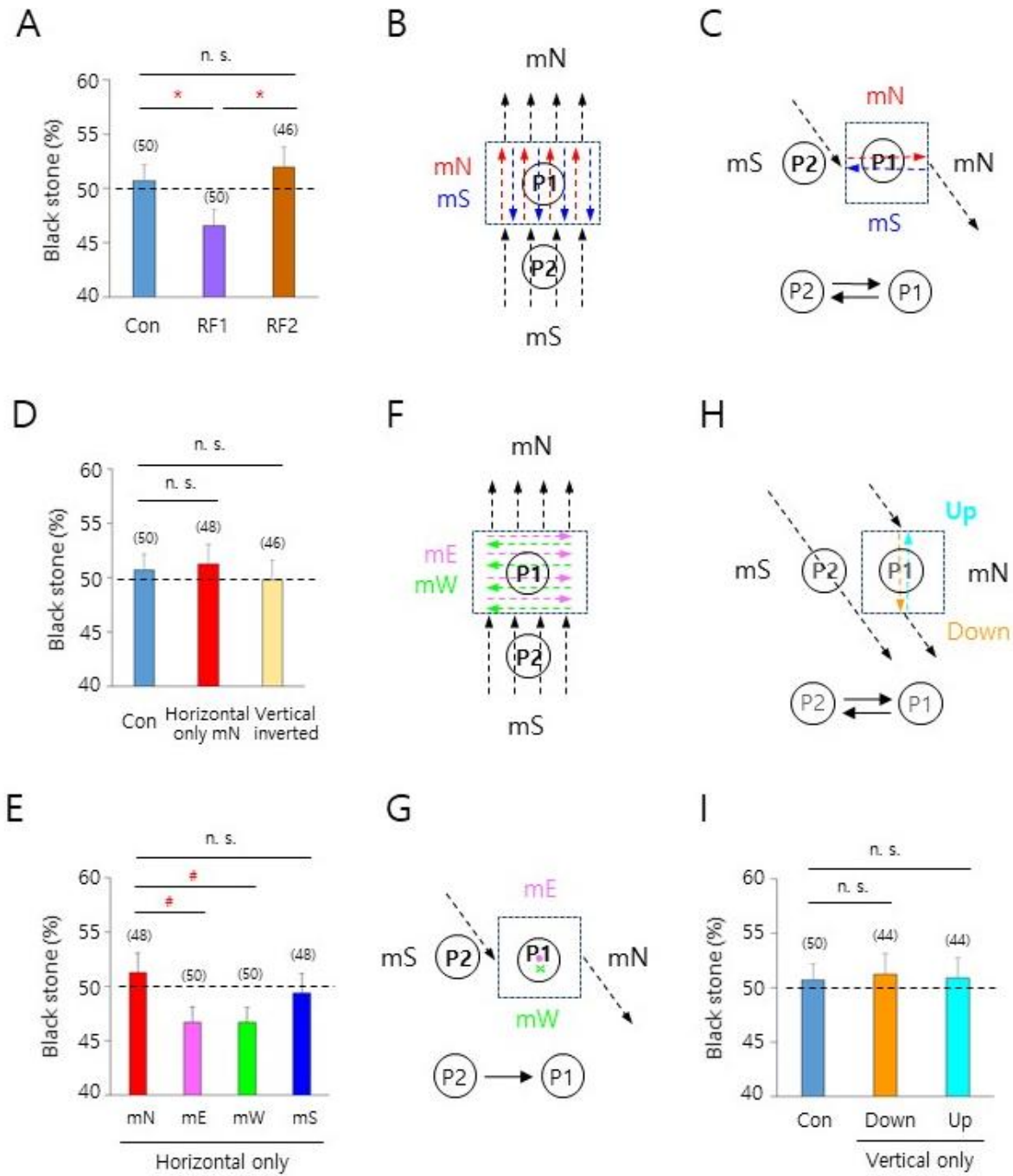

**Fig. 4. The continuum of GMF lines between two persons is crucial to GMF-modulated probabilistic abstract decision-making.**

(**A**) A significant decrease in the rate by the resonance frequency 1 (RF 1: 1.260 MHz) was detected compared to the Con, but not by RF 2 (1.890 MHz). Con, control (the ambient GMF); *, $P < 0.05$ by a two-sample *t*-test. (**B** and **C**) A schematic drawing of the relative position of P1 and P2, magnetic field lines of the ambient GMF or modulated magnetic field at the location of P1's head (*i.e.*, the core of the Helmholtz coils), and the direction of influence flow in the GMF-modulated decision-making between P1 and P2. (**D**) There is no notable difference between the rates of the control and horizontal component alone (horizontal only mN) directed toward magnetic north (mN) or the vertical component-inverted ambient GMF (vertical inverted). (**E**) The comparison of the rate between the control (mN) and mE, mW, or mS, which were the horizontal component only magnetic fields oriented toward the magnetic east, west, and south, respectively. #, $P < 0.025$ by the percentile bootstrap analysis. (**F** to **H**) The relative position of P1 and P2. (**I**) There was no notable difference between the rates of the control (the ambient GMF) and the vertical component only of the GMF oriented downward or upward. The total intensity of each of magnetic fields was 45 µT (A, D, E, and I). n. s., not significant by a two-sample *t*-test; horizontal dashed lines, 50% theoretical probability for black stone rate; error bars, SEM. The number of subjects for each group is indicated above bar graphs. Top views (B) and (F), east side views (upper) and the affecting directions (lower) (C, G, and H). Black mN and mS, the direction of the ambient magnetic north and south, respectively; red mN, blue mS, pink mE, green mW, sky-blue Up and golden Down, the different GMF-modulated experimental conditions at the P1 location; P1, the north seat player; P2, the south seat player; circles, locations of the players; dashed rectangles, the contour of the vertical axis (B and F) and east-west axis (C, G, and H) of the Helmholtz coils; different colors- and black-dashed arrows, a conceptual representation of magnetic field lines of the markedly modulated magnetic fields at the location of P1's head and the marginally changed ones at the location of P2's head or ambient GMF, respectively; arrow tip (●) and tail (×) at the location of P1's head (G), a conceptual representation of magnetic field directions; black arrows, the influence directions between P1 and P2.

## The magnetic interplay between players influences abstract decision-making

In line with the results of the near-zero GMF in Fig. 3E and RF1 in Fig. S3E, the same stepwise analysis showed that the mE condition significantly reduced the rate of the odd cases of the step 1, a primary influencing target, compared to the horizontal only mN control (Fig. 5A and Data S8). These results raised the question of whether the reduced rates of the north seat players under the near-zero GMF, RF1, mE, and mW were caused by merely the north seat players' abnormal responsiveness to those magnetic conditions or a possible aberrance in potential magnetic interplay between the north and the south seat players possibly through the continuum of magnetic fields. To determine which one is the case, different wavelengths of light were exposed to the eyes of *south* seat players since a range of wavelengths of light are known to impact on magnetic perception (*36, 37*). The black stone rate of the north seat players was significantly increased by > 500 nm light containing no blue light, but not by > 400 nm light or Con (g) goggles that penetrate any wavelength of light, suggesting that no magnetic response of the south seat players under > 500 nm light (*36, 37*) might provide a remarkable winning advantage over the north seat players in the zero-sum game (Fig. 5B and Data S8). Consistently, the same stepwise analysis showed that the > 500 nm light condition notably recovered the rate of the odd cases in the step 1, in contrast to the mE condition in Fig. 5A, and further significantly enhanced the rate of even cases in the step 1, compared to the horizontal only mN control (Fig. 5C and Data S8). Importantly, the contrasting black stone winning rates of the odd cases in the step 1 appeared from the beginning in the 2/3 of earlier trials for mE and nearly entire trials for > 500 nm light conditions, compared to the mN control (Fig. 5, D–F, and Data S8). The combined effects of distinct changes by mE and > 500 nm light contributed to the significant results of mE (Fig. 5A) and > 500 (Fig. 5, B, and C), respectively, although each of the changes in the trials was not remarkable. These results highlight that trial-based magnetosensitive prompt mutual interactions between the players are crucial to obtain a higher black stone rate in decision-making. Taken together, the results suggest a conceptual model in which the GMF-responsive mutual influences between two people through the continuum of GMF lines modulate their probabilistic decision-making (Fig. 5G). In this model, local geomagnetic anomalies and the distances between players were not tested or included to simplify the conditions for reaching to conclusions. Anyone at different inclinations or declinations in the model with an orientation maintaining the continuum of GMF lines with another person, seems to be advantageous in probabilistic decision-making with sensing GMF, if (s)he is in the state of GMF-sensitive (*e.g.*, appropriate blood glucose level).

**Discussion**

This study reveals convincing evidence for the implication of GMF in probabilistic *abstract* decision-making in humans. Realistic pressure on optimal choice always exists under the usual circumstances with insufficient information for decision-making (*46–48*). In the post-experiment questionnaire, the subjects answered that they did not perceive magnetic fields or experience any strange feelings during the game (approximately 98%). The unconscious GMF-modulated decision-making framework may give humans a seminal advantage over the ordinary sensory modalities (*e.g.*, the five senses)-based framework. Indeed, besides the experimental results in the laboratory, the large-scale empirical data from the *bona fide* professional Go matches and *in situ* stone selection games support this possibility, wherein the players carried out the matches or games under random feeding and the ambient geomagnetic conditions including magnetic directions at the locations, as modeled in Fig. 5G. An exception in the model, for example, would be the person 5 in a partnership with another one at a location on the geomagnetic equator region. If the GMF lines passing through the person 5's eyes are perpendicular to the ambient GMF lines because of the severe anomaly of the local GMF by natural or man-made structures, the person 5 would not be capable of influencing another person to achieve higher probability for oneself in a decision-making due to the lack of the continuum of GMF lines between them. Other exceptions may occur at any locations on the globe if such a particular condition is applied. Additionally, personal sensibility to the GMF (*37*) may further affect individual feasibility of the magnetic decision-making framework.

In a retrospect, occasionally encountered apparent discrepancies between theoretical and empirical probabilities have usually been explained by the law of large numbers (*50*) or deterministic mechanisms (*14–16*). The present study demonstrates that the GMF-modulated probabilistic decision-making in humans might have been a hitherto unknown cause of these discrepancies. As a paradox against the theoretically even chance proposed by quantum phenomenon (*17, 18*), the novel spin chemistry-based magnetic field resonance (*38–40, 51*) mechanism revealed in the present study suggests that the GMF-mediated modulation might be a channel for affecting abstract decision-making as an information flow from the quantum to the classical world, manifesting theoretical probability into empirical probability. Nevertheless, the reason why the odd cases but not even cases in the step 1 were the primary influential target in the abstract decision-making remains to be obscure, but it could be inferred as an unconscious selectivity by an unknown mechanism or just an unconscious bias.

Our results indicate that a decision-making does not rely on magnetic directional information from conventional polarity or inclination compass, in which both can directly influence actual magnetic movements (*19–21*). Instead, magnetic interplay (inward/outward influence) between persons is crucial to abstract decision-making in the novel

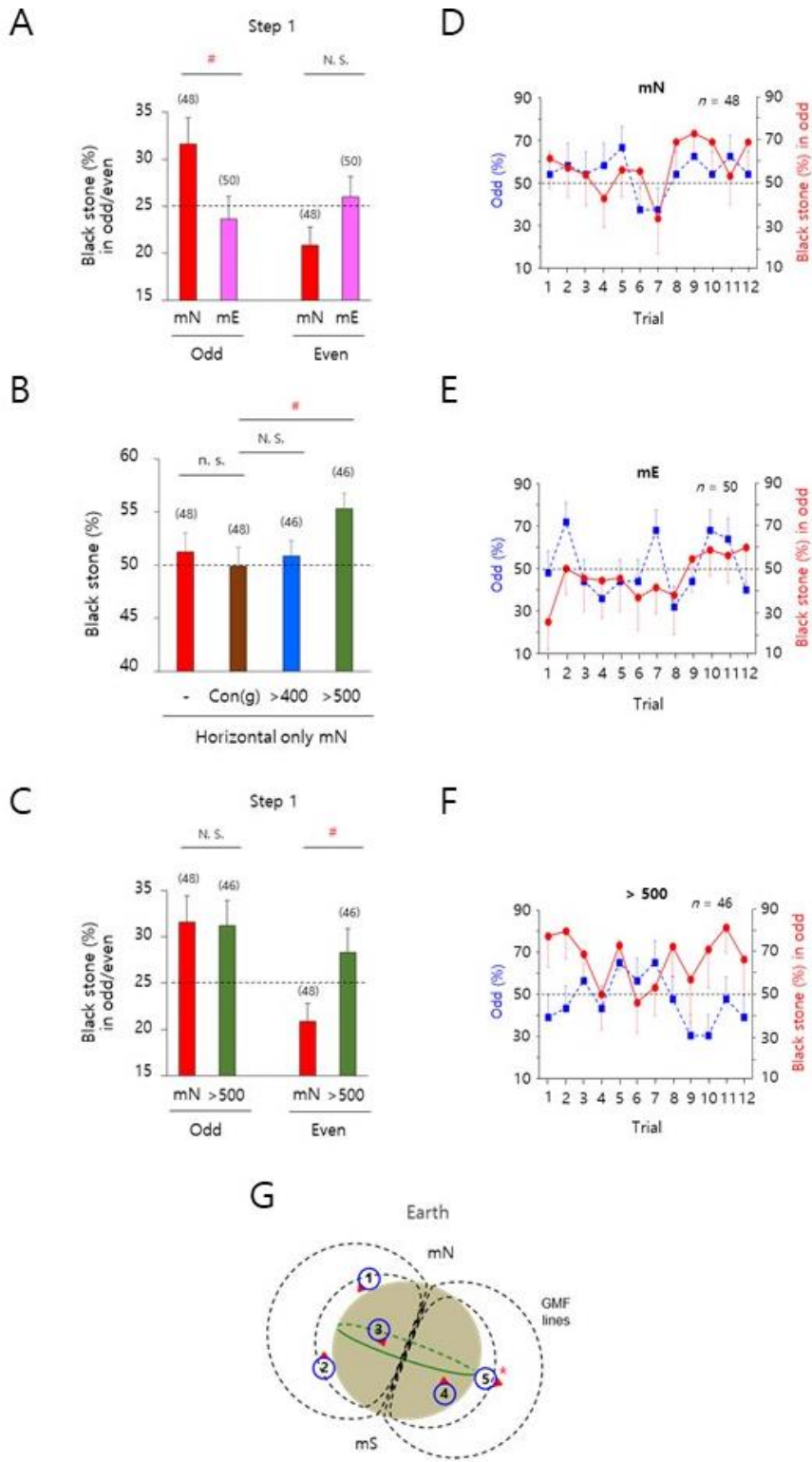

**Fig. 5. Magnetic interplay between two persons influences the GMF-modulated decision-making.**

(**A** and **C**) The black stone rates of the north seat players by a stepwise analysis. The same data of 1–12th trials of mE in Fig. 4E and > 500 (B) were differently analyzed by odd/even in the step 1. Note that each of the distinct changes by mE and > 500, compared to the mN, substantially contributed to the significant results of mE in Fig 4E and > 500 (B), respectively. mN and mE, the horizontal component only magnetic fields oriented toward the magnetic north and east, respectively; > 500, the filter goggles transmitting light higher than 500 nm. (**B**) The comparison of the rates in the horizontal only mN, Con (g), > 400 and > 500. Con (g), the goggles control, not filtering any wavelength of light; > 400, the filter goggles transmitting higher than 400 nm light. #, $P < 0.025$ (A), $P > 0.975$ (B) and $P > 0.975$ (C), by the percentile bootstrap analysis; n. s. and N.S., not significant by a two-sample *t*-test and the percentile bootstrap analysis, respectively; horizontal dashed lines, 25% or 50% theoretical probability for each y-axis index of the graphs; error bars, SEM. The number of subjects for each group is indicated above bar graphs. (**D** to **F**) A time series analysis of the same data of 1–12th trials of horizontal only mN and mE in Fig. 4E and > 500 (B). The blue squares (■) and red circles (●) indicate the odd rate and black stone winning rate in the odd cases, respectively, of the north seat players in the step 1 by trial. In the mN condition (D), an odd rate was frequently higher over the theoretically expected rate (50%) than the conditions mE (E) and > 500 (F). Additionally, the trend of continued low (E) or high (F) level in the black stone rate of the odd cases was detected from the beginning of the trials, compared to the control (D). The pattern showed that the rates of seven trials in the first eight trials (88%) were lower (E) and the rates of 10 trials out of the entire trials (83%) were higher (F) than the theoretically expected rate, respectively. Horizontal dashed lines, 50% theoretical probability for odd rate or black stone rate in the odd case; error bars, SEM. The number of subjects for each graph is indicated at the upper right corner. (**G**) A schematic model shows how GMF continua possibly modulate probabilistic decision-making in humans. mN and mS, the Earth's magnetic north and south, respectively; black-dashed lines, GMF lines; green ellipsoidal solid/dotted line, the geomagnetic equator region; 1–5, the locations of model persons; blue circles, persons' heads; red triangles on the heads, the sight directions of each person. *, an exceptional case that is described in the discussion.

magnetic field resonance-dependent mechanism. Indeed, the findings here may be the first attentive results, supporting potential magnetic interactions between magnetoresponsive individual organisms. It could be speculated that humans are not confined by sensed magnetic information including magnetic polarity or inclination, but exploit high degree of freedom to utilize the information in 3D spaces for abstract decision-making, as theoretically suggested for humans (*52*) and experimentally proven in mice (*53*). Intriguingly, an anisotropic directionality between persons exists regarding the influence flow, wherein the continuum of GMF lines across persons is essential to the outward but not inward influence (Figs. 4 and 5), although its fundamental principle is not well understood. These findings provoke numerous questions to explore, such as 1) What is the inward-outward mechanism for information flow in a person for decision-makings? 2) Is the GMF-modulated abstract decision-making effective over a longer distance between persons? 3) Is a person affected by GMF in abstract decision-making in the absence of another person engaged? 4) Does GMF broadly impact abstract decision-making in humans, which is not limited to probabilistic choices?

In addition to the previous reports on men's magnetic responses to the GMF (*36, 37*), this study demonstrates for the first time that both men and women can sense the GMF to present magnetic behavioral responses. Intriguingly, men and women were influenced by GMF in decision-making at the same restricted blood glucose level (mean, 5.1 mM), despite the different fasting durations. Given the positive correlation between blood glucose level and generated $O_2^{\cdot-}$ concentration in the rat retina (*54*), this result supports the FADH$^{\cdot}$/ $O_2^{\cdot-}$ radical scavenging system in the radical pair hypothesis, which emphasizes the crucial role of $O_2^{•-}$ in magnetoreception at the optimal range of concentration (*37, 55*). This scenario suggests the enhanced sensitivity of magnetic field in the cryptochrome-based radical pair in the eyes (*55–57*) at the restricted blood glucose level. Interestingly, birds can efficiently reduce the level of reactive oxygen species during migration through the endogenous antioxidant mechanisms (*58*), even though they normally maintain 2 to 4 fold higher level of blood glucose compared to mammals in equivalent body mass (*59*). Therefore, the optimum level of $O_2^{•-}$ might be more critical than blood glucose level for magnetoreception, which may be different depending on the species due to subtle differences in the sensing moiety of putative magnetoreceptor molecules (*60*). These possibilities can be tested using *in vitro* human cell systems; for example, the levels of magnetic field-dependent cellular autofluorescence (*61*) can be measured at different concentrations of glucose and $O_2^{•-}$. In the magnetoreceptive organs of animals including migratory birds, *in situ* measurement of those levels can also be conducted. Overall, the findings from the present study may provide unique insights into the function of magnetic sensing in 3-dimensional spaces and the newly identified mechanistic entity to manifest the theoretical probability in the real world. The potential principle and mechanism proposed here may actually impose on other binary decision-makings including coin toss or penalty kick direction in soccer. The potential discrepancy between theoretical and empirical probability from these events in the past or future could be assessed using enormous empirical data. Given the relatively low chance of having the absence of the continuum of GMF lines as shown in Fig. 5G, we need to carefully evaluate the possible anomaly of GMF at the event locations, *i.e.*, potential anisotropy in magnetic influence between stakeholders and appropriate controls for interpreting data correctly.

This present study could be a corner-stone for identifying mysterious forces in decision-making, *e.g.*, GMF. The ever-existing "magnetic communication field" may not only converge empirical probability into theoretical probability in most cases, but also ironically can produce unequal empirical probability under particular conditions in some practical decision-making cases.

## Materials and Methods

### *Stone selection in professional Go matches*

All the final-round professional Go matches were held in either a quiet room or a hall. The stone selection was conducted according to the rules of the Korea Baduk Association (*49*). At the beginning of a match, two players sat facing each other across the board and performed the stone choice process (see Movie S1 (*62*)). In the step 1, the player 1 who has been a professional longer than the other player (player 2), grabs a handful of white stones as many as (s)he wants in his or her right hand and then places the hand on the board to cover the stones. The number of white stones is usually greater than 10. In the step 2, the player 2 places one or two black stone(s) on the board. To determine who would take black or white stones for the match, the player 1 reveals whether the number of the white stones is odd or even by aligning the white stones as separate sets comprised of two or four stones. If both white and black stones turned out to be matched either odd or even (*i.e.*, the "failure" of player 1 and "success" of player 2), each player keeps the stone color (s)he already has. However, if the two sets of stones were mismatched: odd–even or even–odd (*i.e.*, the success of player 1 and the failure of player 2), players traded the stone buckets.

### *The analysis of stone choice data from professional Go matches*

Data from professional Go matches supervised by the Korea Baduk Association were obtained through the official request by the corresponding author. Data elements include the date and location of matches, the title of the championship, the status of the matches (*e.g.*, preliminary/final-round or special match), stone color for players, names of winners/losers, and the compensation points given to the white stone player. The locations were at many different places in the country, despite some matches were held outside of the country (*e.g.*, China or Japan). A substantial portion of preliminary-round matches with insufficient information and some final-round matches between players who were unaffiliated with the Korea Baduk Association, were not included in the dataset. Thus, the analyzed dataset includes 21,212 matches out of 73,730 total matches (28.8%) (Table S1). The analyzed data include 311 professional Go players (262 men; 49 women) listed in the player ranking, issued as of July 2018, which was the very first ranking after February 2018. In a separate analysis, the dataset comprised the player ranking as of July 2018 and the black stone selection rate for each player in every year between 2009 and 2017; players who had fewer than eight matches (approximately one-third of the average number of matches per player annually) in the corresponding year were excluded. Specific numbers for the data used in each analysis are indicated in the text or figure legends.

*In situ stone choice games*

*In situ* stone choice games were executed at 16 different locations distributed over the country in either quiet classrooms (Department of Baduk Studies Myongji University or Korea Baduk middle and high school) or training rooms (private amateur Go training academies) between November 2018 and September 2020. These were performed in the cooperation with the institutions at the appointed time under the supervision of an experimenter—volunteered 293 participants (220 men; 73 women) of 16 groups (18.3 ± 7.4 people/group, mean ± SD). The participants and an experimenter wore masks to reduce COVID-19 infections during the games in the middle of the pandemic (March 2020–September 2020). Before the beginning of games, all the participants were informed of the aim and procedure of the games, the financial compensation for the participation, and the additional reward for the selection of black stone. These stone choice games were conducted by the same rule as the stone selection in professional Go matches, with some exceptions. 1) Two participants in each group were randomly chosen for the first game. In the step 1, a participant with higher master level ("Dan," 0–9 grade) was player 1, and the other participant with lower level was player 2 in the match. In cases of the same Dan for the participants, the elderly participant would be the player 1. 2) The match comprised two sets and 20 trials of stone choice per set. Using the same game rules, both players took a turn to grab white stones in the trials in a set—player 1 grabbed white stones in the first trial, and player 2 grabbed white stones in the second trial for the stone choice (see Data S9 for record form and an example; Dan was included in the participant's information for *in situ* stone choice games). Participants were instructed to grab at least 10 white stones in the step 1 (if it was less than 10, the trial was nullified and retried) and not to use a predetermined pattern but to decide the number of black stones (one or two) extemporaneously in the step 2 during the entire game. The game information was determined and then recorded on the record form for each trial, set, and game by the participants themselves, with both players' confirmation under the supervision of an experimenter (the correction rate of judgment error in the game information, approximately 2%). 3) The players switched seats, "player 1" and "player 2" each other, before starting the second set in a game (Fig. 1A); the same game rule and procedure in 2) were applied to the second set. 4) The black stone rate (%) of a player in a game was calculated as the number of trials with success for black stone/total number of trials (*i.e.*, 40) × 100. 5) All the participants played the second game with another random opponent player within a group, and a player's black stone rate in the analyzed data was the mean of the two games of a player.

## *Subjects for stone choice experiments in the laboratory*

Participants are composed of 55 men (age, 21–27 years; mean, 23 years; body mass index, 20–34 kg/m$^2$; mean, 24 kg/m$^2$) and 53 women (19–26 years, mean 21 years; body mass index, 15–30 kg/m$^2$, mean 21 kg/m$^2$) volunteers without reported physical disabilities or mental disorders including color blindness and claustrophobia (*36, 37*). All the subjects were undergraduate students who are amateur Go players, and thus they were not categorized by the level (Dan). They were informed of the objective of the study, experimental procedures, and financial compensation for participation and were asked to follow the rules of the study. To motivate the subjects to win the game and achieve a higher black stone rate, two kinds of financial rewards were provided individually—one for winning a game and the other for higher monthly black stone selection rate. Before each experiment, subjects underwent short-term fast for either ~ 7 h (around 9:00 am–4:00 pm or 11:00 am–6:00 pm) or ~ 20 h (around 2:00 pm–10:00 am or 6:00 pm–2:00 pm), depending on the fast duration and time point for experiments. Neither food nor medical treatments except pure water was allowed during the fast between the last meal and test (*36, 37*). To exclude the insufficient sleep effect, normal night sleep that is at least 6 h between 10 pm and 8 am on the test day, was required (*36, 37*). Before starting each experiment, the subjects were stabilized on a chair for approximately 15 min in a waiting room next to the testing room. In particular, the subjects were supervised to remove any colored lens and detachable metallic or electromagnetic items including coins, watches, glasses, earrings, hairpins, and mobile phones from their bodies. Based on the assessments of a pre-experiment questionnaires and blood glucose levels on the predetermined subjects before starting the first game (see 'Stone choice experiments in the laboratory' below), any subjects who had not followed these rules were rescheduled. Some subjects were excluded from several experiments for personal reasons, such as conflicting schedule or COVID-19-related symptoms. The study was approved in advance by the Institutional Review Board of Kyungpook National University (KNU-2021-0153). All procedures followed the regulations for human subject research, and informed consent was obtained from all the subjects.

## *GMF modulation and magnetic fields oscillation*

The ambient GMF in the core of the Helmholtz coils in a testing room had the total intensity of 45.0 μT, inclination of 53°, and declination of −8° (Daegu City, Republic of Korea), which were maintained in the laboratory throughout the period of stone choice experiments. The testing room was shielded by a rectangular parallelepiped Faraday cage comprising 10 mm thick aluminum plates, and grounded during the entire experiment (*36, 37*). To provide the subjects' eyes with the indicated GMF-like magnetic fields, the same coil system from our previous studies (*23, 24, 36, 37*) was applied to modulate the total intensity, inclination, and the direction of the magnetic field, depending on the experimental conditions (Table S2). Briefly, it comprised three double-wrapped, orthogonal, and rectangular Helmholtz coils (1.89 × 1.89 m, 1.89 × 1.80 m, and 1.98 × 1.98 m for the north-south, east-west, and vertical axes, respectively) electrically-grounded with copper mesh shielding. A player sat on a non-metallic homemade chair at the center of the Helmholtz coils (north seat) with his head positioned in the middle space of the vertical axis of the coils, and the other player sat on the same type of chair outside of the coils (south seat) on the ambient magnetic north-south axis (Fig. 2A and Fig. S2, B, and C). The two players sat facing each other at the distance of 1.30 m across the board of 60 × 70 × 72 cm (L × W × H), and an experimenter sat on a non-metallic chair outside of the coils at the distance of 1.35 m east from the two players to conduct the experiments. The modulated GMF at the glabella of subjects on the north seat varied markedly as indicated, but the accompanying change in the GMF at the same region of subjects on the south seat was relatively marginal (Table S2). The field homogeneity at the position of the subject's head was approximately 95%, as measured using a 3-axis magnetometer (MGM 3AXIS; ALPHALAB, USA) (*36*, *37*). The oscillating magnetic fields, RF 1 and RF 2 (1.260 and 1.890 MHz, respectively; mean 100 nT for each) were exposed vertically (37° relative to the ambient GMF) to the north seat players during the corresponding set of the game (approximately 13 min) (*37*). To produce the oscillating magnetic fields, the same system comprising a function generator, amplifier, and calibrated coil antenna, was used as reported in the previous study (*37*). The measurement of oscillating magnetic fields was performed on the glabella of the subjects using the same spectrum analyzer with the calibrated loop antenna and magnetometer (the band widths of the RF 1 and RF 2 were 0.020 and 0.019 MHz ("average," √3 kHz), respectively, at the bottoms of the peaks), as in the previous study. The electromagnetic noise in the cage including the switch button module for GMF modulation and the antenna for generating oscillating magnetic fields was measured and maintained constantly (see ref. *37*). The temperature at the position of the subjects' head was maintained at 25 ± 0.5°C (Data logger 98581; MIC Meter Industrial, Taiwan) (*36*, *37*).

*Stone choice experiments in the laboratory*

To habituate the subjects to the stone choice game, all the subjects participated in two preliminary games between two randomly matched opponents by the same procedures of the '*in situ* stone choice games' under the supervision of the experimenters in the waiting room. The results of these preliminary games are not included in the data analysis. The stone choice experiments in the testing room were conducted according to the same rules in the preliminary games with some differences. 1) Experiments were performed within 90 min window at 10:00–11:30 am, 2:00–3:30 pm, 4:00–5:30 pm or 6:00–7:30 pm (local time, UTC + 09:00) (time spent for an experiment: 50–70 min; mean ≈ 1 h). The subjects underwent different durations of fast depending on the experiment, as described above. If the determined blood glucose levels of any of the two players before the first game varied by more than 15% relative to the mean (*37*), the experiment was postponed to a later date (approximately 2% of experiments). 2) In the experiment, two randomly predetermined subjects belonging to a subject group of the same gender (men or women) were tested in two consecutive games under different experimental conditions. The two subjects were randomly assigned as a player 1 or 2 in the first game by an experimenter and then assigned as the other player in the second game. Before the experiments, the subjects were asked to sit facing toward the opponent player throughout the games, while they were allowed to turn their head or eyes toward the experimenter to see hand cues for the game progress or the scoreboard and any directions to refresh their mind for decision-making. Either the transparent acrylic or opaque double-sided hardboard acrylic screen (0.3 × 60 × 50, 2 × 60 × 90 cm, respectively; L × W × H; Fig. S2, B, and C) stood on the board between the players, depending on the games. In the "opaque screen" games, both players had earplugs and earmuffs to prevent auditory cues. Subjects at both seats were illuminated by light from a diffused light-emitting diode (*37*) during games with a transparent or opaque screen (290 or 280 lux (lx) on the glabella, respectively). In some games with the opaque screen, non-filtered/filtered light was provided to the *south* seat players using homemade filter goggles (*36, 37*) that were worn throughout the games (non-filtered goggles, 130 lx; > 400 nm, 108 lx; > 500 nm, 72 lx on the glabella). Before the first game in the experiment, both players were reminded of the two kinds of financial rewards described above. 3) The same procedure as 2) in the "*in situ* stone choice games*"* was conducted by an experimenter's strict step-by-step hand cues (see Movie S2). The experimenter judged and recorded the game information for each trial and set on the record form. A non-metallic scoreboard (Fig. S2A) stood in front of an experimenter and was manually maintained by the experimenter, so that a real-time score could be provided to the players. The experimenter confirmed the correctness of the written information on the record form at the end of the set. 4) The players switched the seats for the second set of games under the guidance of an experimenter, and the same procedure in 3) was carried out in the second set. 5) After the second set of information was validated, the black stone rate (%) for each player in a game was calculated as the number of trials with success for black stone/total number of trials (*i.e.*, 40) × 100 (A correction rate for error in a game is approximately 1%). The game score was posted on a scoreboard for players to see. 6) According to the procedures 3)–5) above, the two players participated in the second game under different experimental conditions by the supervision of an experimenter. After the completion of the experiment, the subjects were asked to separate in the waiting room and then the post-experiment questionnaire was conducted based on the recorded game information. The questions cover what they experienced during the experiment including whether they could perceive magnetic fields or experience any strange feelings at any moments during the game and intentionally grab white stones to be odd or even in the step 1. Experiments under different conditions were dispersed and performed in a random order with an interval of at least 3 days for the same subject between experiments. All experiments were performed in a double-blinded manner. The experimenters who conducted the stone choice games were aware of the experimental conditions including the fast duration of the subject, the type of screen (transparent or opaque), and whether the subject was wearing filter or non-filter goggles. Another experimenter, who analyzed the data, was not informed of the experimental conditions. Thus, none of the experimenters was aware of all the information including the subjects, experimental conditions, collected data, and process of data analysis.

*Statistical analysis*

To determine the significance of data, linear fit analysis, a two-sample *t*-test (*37*) or the percentile bootstrap method (*37*) was applied. The linear fit analyses of stone selection data from professional Go matches and *in situ* stone choice games were performed using the software Origin 2019 (OriginLab, Northampton, USA). The analysis of the *north* seat players' data from the stone choice experiments in the laboratory was performed as below. The black stone rate (%) for the north seat player in a game was calculated as the number of trials with success for black stone/total number of trials at the north seat (*i.e.*, 20) × 100. For the time series analysis by trial, the black stone rates of all subjects were averaged for each trial in the 1–20th trials. The detailed calculation formulas for the time series and stepwise analyses are described in Materials and Methods S1. To verify the suitability of the *t*-test, each of the group datasets was examined using the Anderson–Darling test to determine whether the data showed a normal distribution (*37*) (Data S10). To determine if the difference between the means of the two data sets was significant, the two-sample *t*-test was used when the two data sets followed a normal distribution (Data S11), and the percentile bootstrap method (*37, 63*) was employed (95% confidence interval) if any of them deviated from a normal distribution (Data S12, and Data S13 for raw data). To evaluate the blood glucose level, the two-sample *t*-test or percentile bootstrap method was adopted, depending on the results of the Anderson–Darling test, as described above. Statistical values are presented as the mean ± standard error of the mean (SEM). n.s., not significant by a two-sample *t*-test; N.S., not significant by the percentile bootstrap analysis. *P*-value, $< 0.05$ by linear fit analysis; *, $< 0.05$ or **, $< 0.01$ by a two-sample *t*-test; #, $< 0.025$ or $> 0.975$ by the percentile bootstrap analysis, were regarded as significant.

### Acknowledgements

We are grateful to Korea Baduk Association for the valuable data of professional Go matches and Department of Baduk Studies (Myongji University), Korea Baduk middle and high school, many Go training academies for help in the *in situ* stone choice games. We thank students Seungbok Yang and Hyebin Kim for technical assistance in calculation of the raw data and Prof. Hongjoo Kim and Dr. JooYoung Lee (Kyungpook National University) for help to format the manuscript.

### Funding

Basic Science Research Program (2021R1A2C1094765) to KSC through the NRF funded by the Korea government (MSIT).

### Author contributions

Conceptualization: K.-S.C.
Methodology: K.-S.C., S.-C.K. and Y.K.
Resources: K.-S.C. and S.H.J.
Investigation: K.-S.C. and I.-T.O.
Data analysis: K.-S.C., I.-T.O., Y.K. and Y.-H.K.
Visualization: K.-S.C. and I.-T.O.
Funding acquisition: K.-S.C.
Project administration: K.-S.C.
Supervision: K.-S.C.
Writing–original draft: K.-S.C.
Writing–review & editing: Y.-H.K., K.-S.C., I.-T.O., Y.K., S.H.J. and S.-C.K.

### Competing interests

Authors declare that they have no competing interests.

### Data and materials availability

All data are available in the main text or the supporting information including the raw data (https://github.com/kschae5/arXiv).

### References

1. McGarry, T. & Franks, I. M. On winning the penalty shoot-out in soccer. *J. Sports. Sci.* **18**, 401-409 (2000).
2. Hahn, U. & Warren, P. A. Perceptions of randomness: why three heads are better than four. *Psychol. Rev.* **116**, 454-461 (2009).
3. Wenzel, K., Schindler, S. & Reinhard, M. A. General belief in a just world is positively associated with dishonest behavior. *Front. Psychol.* **8**, 1770 (2017).
4. Piper, S. K. et al. Exact replication: foundation of science or game of chance? *PLoS. Biol.* **17**, e3000188 (2019).
5. Nash, J. Non-cooperative games. *Ann. Math*. **54**, 286-295 (1951).

6. Koller, D. & Megiddo, N. The complexity of two-person zero-sum games in extensive form. *Games. Econ. Behav.* **4**, 528-552 (1992).
7. Davies, J. The rules and elements of Go. *Ishi. Press* (1977).
8. Chaslot, G., Saito, J. T., Bouzy, B., Uiterwijk, J. W. H. M. & van den Herik H. J. Monte-Carlo strategies for computer Go. *Proc. The 18th Belgium-Netherlands Confer. Artf. Intell.* **18**, 83-90 (2006).
9. Benson, D. B. Life in the game of Go. *Infor. Sci.* **10**, 17-29 (1976).
10. Silver, D. et al. Mastering the game of Go with deep neural networks and tree search. *Nature.* **529**, 484-489 (2016).
11. Kim, J. & Kim, H. J. The advantage of playing first in Go. *Chance.* **8**, 22-26 (1995).
12. Kim, J. & Jeong, S. H. Is 6.5 dum optimal? *Proc. The 3rd Int. Confer. Baduk.* **3**, 157-166 (2005).
13. Kim, J. A study on the optimal size of dum in professional. *Kor. J. Appl. Stat.* **20**, 245-255 (2007).
14. Diaconis, P., Holmes, S. P. & Montgomery, R. Dynamical bias in the coin toss. *SIAM. Rev.* **49**, 211-235 (2007).
15. Clark, M. P. A. & Westerberg, B. D. Holiday review. How random is the toss of a coin? *CMAJ.* **181**, E306-E308 (2009).
16. Ferrie, C. & Combes, J. How the result of a single coin toss can turn out to be 100 heads. *Phys. Rev. Lett.* **113**, 120404 (2014).
17. Albrecht, A. & Phillips, D. Origin of probabilities and their application to the multiverse. *Phys. Rev. D* **90**, 123514 (2014).
18. Adler, S. L. Where is quantum theory headed? *J. Phys.: Conf. Ser.* **504**, 012002 (2014).
19. Johnsen, S. & Lohmann, K. J. The physics and neurobiology of magnetoreception. *Nat. Rev. Neurosci.* **6**, 703-712 (2005).
20. Wiltschko, R. & Wiltschko. W. Magnetoreception. *Bioessays.* **28**, 157-168 (2006).
21. Lohmann, K. J., Putman, N. F. & Lohmann C. M. The magnetic map of hatchling loggerhead sea turtles. *Curr. Opin. Neurobiol.* **22**, 336-342 (2012).
22. Guerra, P. A., Gegear, R. J. & Reppert, S. M. A magnetic compass aids monarch butterfly migration. *Nat. Commun.* **5**, 4164 (2014).
23. Bae, J. E. et al. Positive geotactic behaviors induced by geomagnetic field in *Drosophila*. *Mol. Brain.* **9**, 55 (2016).
24. Oh, I. T. et al. Behavioral evidence for geomagnetic imprinting and transgenerational inheritance in fruit flies. *Proc. Natl. Acad. Sci. USA.* **117**, 1216-1222 (2020).
25. Begall, S., Cerveny, J., Neef, J., Vojtech, O. & Burda, H. Magnetic alignment in grazing and resting cattle and deer. *Proc. Natl. Acad. Sci. USA.* **105**, 13451-13455 (2008).
26. Hart, V. et al. Dogs are sensitive to small variations of the Earth's magnetic field. *Front. Zool.* **10**, 80 (2013).
27. Bazalova, O. et al. Cryptochrome 2 mediates directional magnetoreception in cockroaches. *Proc. Natl. Acad. Sci. USA.* **113**, 1660-1665 (2016).
28. Wiltschko, W. & Wiltschko, R. Magnetic compass of European robins. *Science.* **176**, 62-64 (1972).
29. Phillips, J. B. Two magnetoreception pathways in a migratory salamander. *Science.* **233**, 765-767 (1986).
30. Lohmann, K. et al. Magnetic orientation of spiny lobsters in the ocean: experiments with undersea coil systems. *J. Exp. Biol.* **198**, 2041-2048 (1995).
31. Marhold, S. Wiltschko, W. & Burda, H. A magnetic polarity compass for direction finding in a subterranean mammal. *Naturwissenschaften.* **84**, 421–423 (1997).
32. Baker, R. R. Goal orientation by blindfolded humans after long-distance displacement: possible involvement of a magnetic sense. *Science.* **210**, 555-557 (1980).

33. Westby, G. W. & Partridge, K. J. Human homing: still no evidence despite geomagnetic controls. *J. Exp. Biol*. **120**, 325-331 (1986).
34. Mulligan, B. P. & Persinger, M. A. Experimental simulation of the effects of sudden increases in geomagnetic activity upon quantitative measures of human brain activity: validation of correlational studies. *Neurosci. Lett*. **516**, 54-56 (2012).
35. Wang, C. X. et al. Transduction of the geomagnetic field as evidenced from alpha-band activity in the human brain. *eNeuro*. **6**, 0483-18.2019 (2019).
36. Chae, K. S., Oh, I. T., Lee, S. H. & Kim, S. C. Blue light-dependent human magnetoreception in geomagnetic food orientation. *PLoS. One*. **14**, e0211826 (2019).
37. Chae, K. S., Kim, S. C., Kwon, H. J. & Kim, Y. Human magnetic sense is mediated by a light and magnetic field resonance-dependent mechanism. *Sci. Rep*. **12**, 8997 (2022).
38. Schulten, K., Swenberg, C. E. & Weiler, A. A biomagnetic sensory mechanism based on magnetic field modulated coherent electron spin motion. *Z. Phys. Chem*. **111**, 1-5 (1978).
39. Ritz, T., Adem, S. & Schulten, K. A model for photoreceptor-based magnetoreception in birds. *Biophys. J*. **78**, 707-718 (2000).
40. Hore, P. J. & Mouritsen, H. The radical-pair mechanism of magnetoreception. *Annu. Rev. Biophys*. **45**, 299-344 (2016).
41. Thompson, C. L. et al. Expression of the blue-light receptor cryptochrome in the human retina. *Invest. Ophthalmol. Vis. Sci*. **44**, 4515-4521 (2003).
42. Hochstoeger, T. et al. The biophysical, molecular, and anatomical landscape of pigeon CRY4: a candidate light-based quantal magnetosensor. *Sci. Adv*. **6**, eabb9110 (2020).
43. Xu, J. et al. Magnetic sensitivity of cryptochrome 4 from a migratory songbird**.** *Nature*. **594**, 535-540 (2021).
44. Sarimov, R. M., Binhi, V. N. & Milyaev, V. A. The influence of geomagnetic field compensation on human cognitive processes. *Biophysics*. **53**, 433–441. (2008).
45. Zhang, B. et al. Long-term exposure to a hypomagnetic field attenuates adult hippocampal neurogenesis and cognition. *Nat. Commun*. **12**, 1174 (2021).
46. Stevens, M. & Ruxton, G. D. The key role of behaviour in animal camouflage. *Biol. Rev. Camb. Philos. Soc*. **94**, 116-134 (2018).
47. Bogacz, R., Brown, E., Moehlis, J., Holmes, P. & Cohen, J. D. The physics of optimal decision making: a formal analysis of models of performance in two-alternative forced-choice tasks. *Psychol. Rev*. **113**, 700-765 (2006).
48. Kilpatrick, Z. P., Holmes, W. R., Eissa, T. L. & Josić, K. Optimal models of decision-making in dynamic environments. *Curr. Opin. Neurobiol*. **58**, 54-60 (2019).
49. Korea Baduk Association. http://english.baduk.or.kr (2023).
50. Bernoulli, J. Opera, Volume 2. Sumptibus haeredum Cramer et fratrum Philibert (1744).
51. Xie, C. Searching for unity in diversity of animal magnetoreception: from biology to quantum mechanics and back. *Innovation (Camb)*. **3**, 100229 (2022).
52. Yukalov, V. I. & Sornette, D. Decision theory with prospect interference and entanglement. *Theory. Decis*. **70**, 283-328 (2011).
53. Beron, C. C., Neufeld, S. Q., Linderman, S. W. & Sabatini, B. L. Mice exhibit stochastic and efficient action switching during probabilistic decision making. *Proc. Natl. Acad. Sci. U S A*. **119**, e2113961119 (2022).
54. Du Y., Miller, C. M. & Kern, T. S. Hyperglycemia increases mitochondrial superoxide in retina and retinal cells. *Free Radic. Biol. Med*. **35**, 1491-1499 (2003).
55. Kattng, D. R. Radical-pair-based magnetoreception amplified by radical scavenging: resilience to spin relaxation. *J. Phys. Chem. B*. **121**, 10215-10227 (2017).
56. Player, T. C. & Hore, P. J. Viability of superoxide-containing radical pairs as

magnetoreceptors. *J. Chem. Phys.* **151**, 225101 (2019).
57. Deviers, J., Cailliez, F., de la Lande, A. & Kattnig, D. R. Anisotropic magnetic field effects in the re-oxidation of cryptochrome in the presence of scavenger radicals. *J. Chem. Phys.* **156**, 025101 (2022).
58. Gutiérrez, J. S. et al. Oxidative status and metabolic profile in a long-lived bird preparing for extreme endurance migration. *Sci. Rep.* **9**, 17616 (2019).
59. Braun E. J. & Sweazea K. L. Glucose regulation in birds. *Comp. Biochem. Physiol. B* **151**, 1-9 (2008).
60. Zoltowski, B. D. et al. Chemical and structural analysis of a photoactive vertebrate cryptochrome from pigeon. *Proc. Natl. Acad. Sci. U S A.* **116**, 19449-19457 (2019).
61. Ikeya, N. & Woodward, J. R. Cellular autofluorescence is magnetic field sensitive. *Proc. Natl. Acad. Sci. U S A.* **118**, e2018043118 (2021).
62. The 1st Go match between the AlphaGo and Korean Go master Lee Sedol in 2016. https://www.youtube.com/watch?v=CPmWrfwYJME (2023).
63. Efron, B. The jackknife, the bootstrap and other resampling plans. CBMS-NSF regional conference series in applied mathematics, *Monograph*, vol. 38 (SIAM, 1982).

**Supporting information**

**Materials and Methods S1**

*Calculation formulas for the time series and stepwise analyses*

For the stepwise analysis in Fig. 3 and Fig. S3, the raw data of the 1–12th trials in those figures were applied to the calculation formula below. 1) For "black stone (%)" of a subject in the step 1, the number of trials with success for black stone in the step 1/total number of trials for a subject (*i.e.*, 12) × 100; 2) For "black stone (%)" of a subject in the step 2, the number of trials with success for black stone in the step 2/total number of trials for a subject (*i.e.*, 12) × 100; 3) For "odd/even (%)" of a subject in the step 1, the number of trials with odd/even white stones/total number of trials for grabbing white stones in the step 1 (*i.e.*, 6) × 100; 4) For "odd/even (%)" of a subject in the step 2, the number of trials with one or two black stone(s)/total number of trials for grabbing black stone(s) in the step 2 (*i.e.*, 6) × 100; 5) For "black stone (%) in odd/even" of a subject in the step 1, the number of trials with success for black stone in the case of odd or even white stones/total number of trials for grabbing white stones in the step 1 (*i.e.*, 6) × 100; 6) For "black stone (%) in odd/even" of a subject in the step 2, the number of trials with success for black stone in the case of one or two black stone(s)/total number of trials for grabbing black stone(s) in the step 2 (*i.e.*, 6) × 100. For each of the group rates in the analyses above, the rates of all the subjects were averaged. The calculation formulas for the time series analysis in Fig. 5 are as follows: 1) For "odd (%)" for a group in each trial of the step 1, the number of subjects with odd white stones/total number of subjects in the group × 100. 2) For "black stone (%) in odd" for a group in each trial of the step 1, the number of subjects with success for black stone/total number of subjects with odd white stones in the group × 100.

**Supporting Figures**

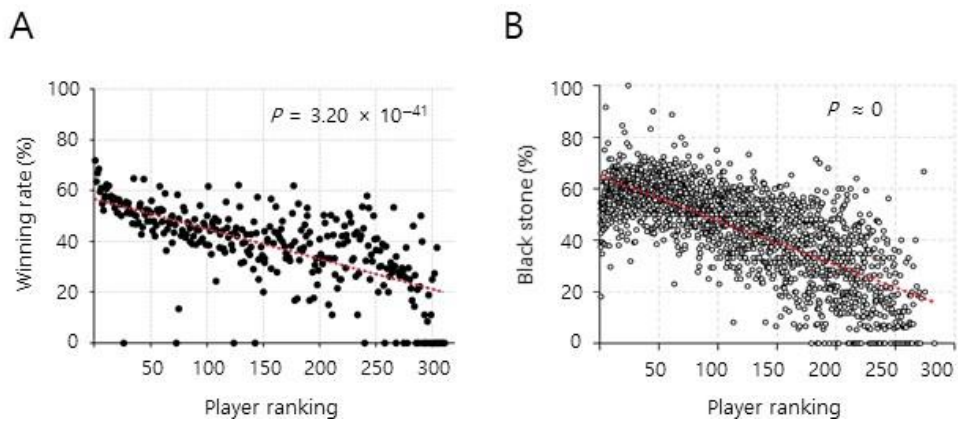

**Fig. S1. Significant inverse correlations between player ranking and a winning rate or a black stone selection rate.**

(**A** and **B**) The winning rate and the percentage of black stone selection were analyzed and displayed in the player ranking among 311 listed players as of June 2018. Linear fit analyses on the data from the final rounds of 21,212 Go matches held during the period mentioned are displayed. Significant inverse correlations between player ranking and winning rate (A) or year-based black stone rate (B) between 2009 and 2017, wherein players who had less than eight matches in a corresponding year were not included. (A) $y = -0.12\,x + 56.60$, $P = 3.20 \times 10^{-41}$ (except outliers ($n = 24$, 7.7%) who had 0% winning rate due to the lack of winning in final-round matches, $P = 5.47 \times 10^{-38}$), Pearson's r $= -0.67$, $n = 311$. (B) $y = -0.17\,x + 64.96$, $P \approx 0$, Pearson's r $= -0.73$, $n = 1861$.

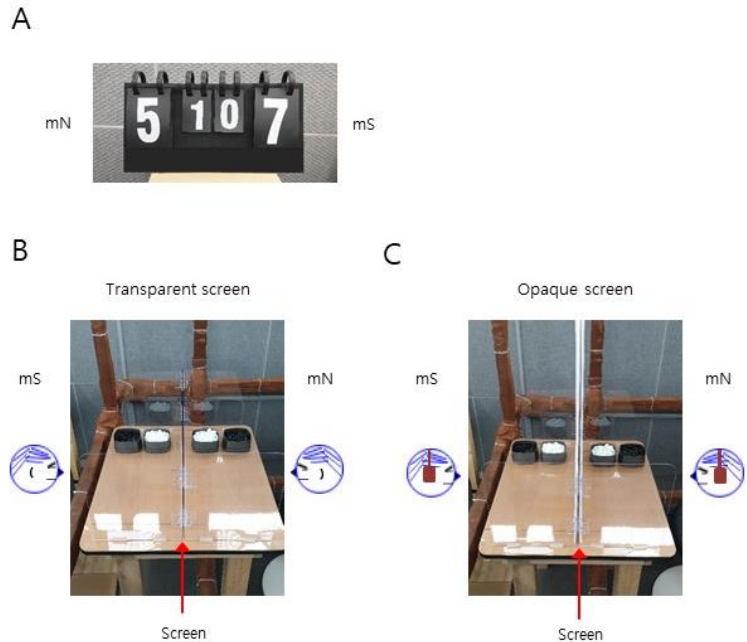

**Fig. S2. The scoreboard and the screens used in the game.**

(**A**) A scoreboard was maintained to post real-time set and game scores to the players by an experimenter. Large numbers indicate the score in a set, while small numbers are game scores for the north and south seat players at a certain time point; mN and mS, the north and south seat direction, respectively. In this example, the game score is "1 to 0" and the real-time score of the ongoing set in the second game is "5 to 7" for the north and south seat player, respectively. (**B** and **C**) An experimenter's views of the transparent or opaque screen on the board between the players are displayed. mN and mS, the ambient magnetic north and south direction, respectively.

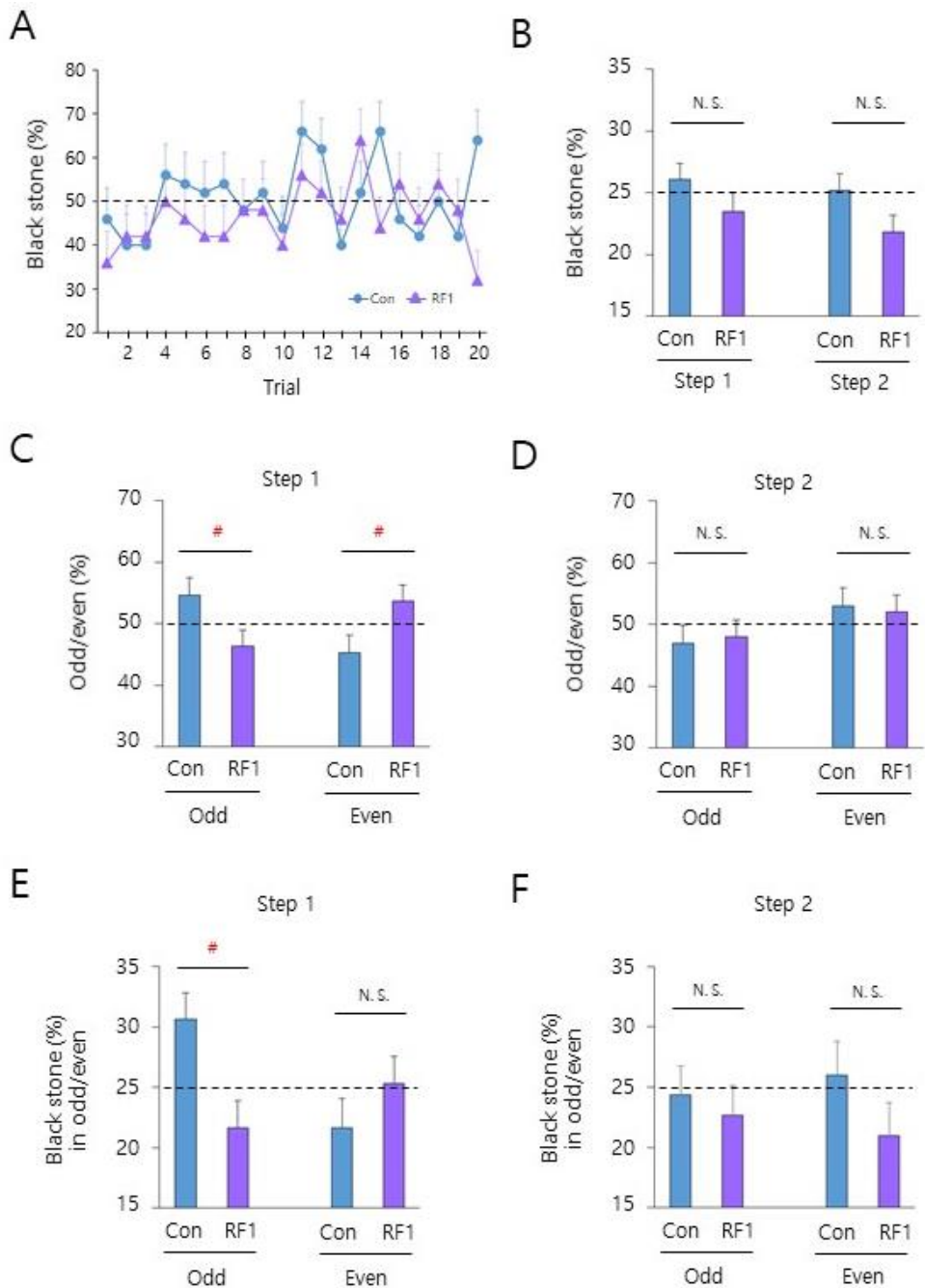

**Fig. S3. A potential magnetic field resonance mechanism underlying the GMF-modulated probabilistic abstract decision-making.**

(A) A profile of the black stone rate of the north seat players in a time series analysis. The same data from the RF1 condition in Fig. 4A was analyzed by trial. Note a continued reduction of the rate by the RF1, compared to the control, up to the 12th trial with a lag at the second and third trials similar to Fig. 3A. (**B** to **F**) The different rates of the north seat players by a stepwise analysis. The data of 1–12th trials were analyzed for the black stone rate, odd/even rate or black stone rate in odd/even by step. Each of the distinct changes by the RF1 in (B to F) contributed to the remarkable results of the profile (A) and the RF1 condition in Fig. 4A. Con, control (the ambient GMF); RF1, the 1.260 MHz resonance frequency; N.S., not significant; #, $P$-value $< 0.025$ in odd case (C), $> 0.975$ in even case (C), and $< 0.025$ (E) by the percentile bootstrap analysis; horizontal dashed lines, 25% or 50% theoretical probability for each y-axis index of the graphs; error bars, SEM. Subjects for each Con or RF1 group in the panels (A to F) are consistently $n = 50$.

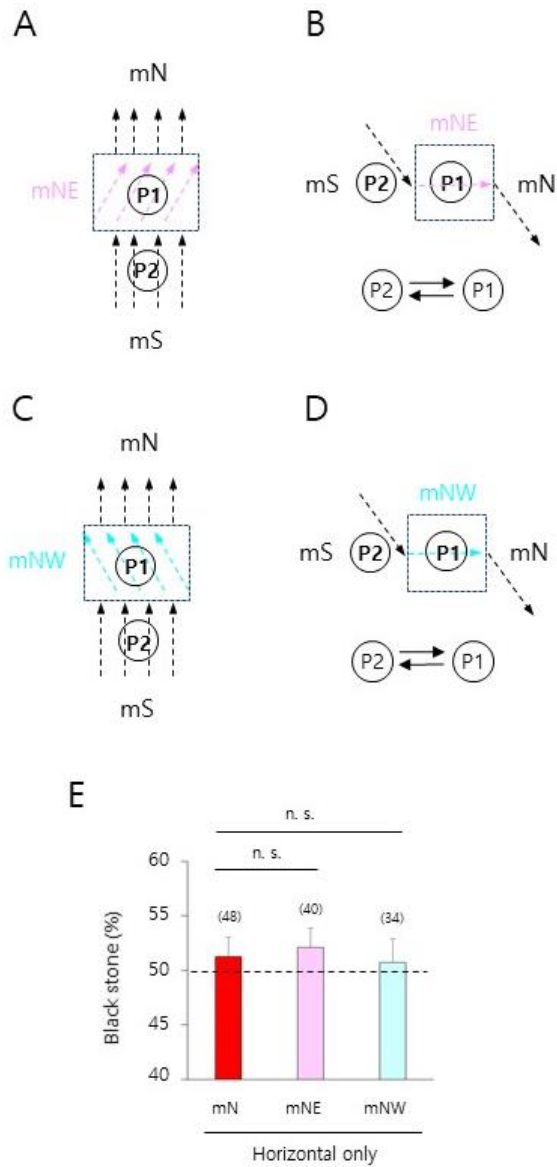

**Fig. S4. A schematic shows the experimental conditions of mNE and mNW, and black stone selection rates under the conditions.**

(**A** to **D**) A schematic display as in Fig. 4. Top views (A and C), east side views (upper) and the influence directions (lower) (B and D). Black mN and mS, the direction of the ambient magnetic north and south, respectively; mNE and mNW, the different GMF-modulated experimental conditions at the P1 location; P1, the north seat player; P2, the south seat player; circles, locations of the players; dashed rectangles, the contour of the vertical axis (A and C) and east-west axis (B and D) of the Helmholtz coils; different colors- and black-dashed arrows, a conceptual representation of magnetic field lines of the markedly modulated magnetic fields at the location of P1's head and the marginally changed ones at the location of P2's head or ambient GMF; black arrows, the influence directions between P1 and P2. Note the directions of different color-dashed arrows at P1's head in (A to D). (**E**) The comparisons of the black stone selection rate between the mN control and mNE or mNW. n.s., not significant by a two-sample *t*-test; horizontal dashed line, 50% theoretical probability for black stone rate; error bars, SEM. The number of subjects for each group is indicated above bar graphs.

**Table S1. Professional Go matches data used for statistical analyses.**

| Year | Number of matches | | | Number of matches by compensation points [a] | | | | |
|------|---|---|---|---|---|---|---|---|
| | Final round [b] | Preliminary round | Total | 5.5 | 6.5 | 7.5 | 8 | 0 [c] |
| 2000 | 520 | 2480 | 3000 | 749 | 2039 | 5 | 14 | 193 |
| 2001 | 533 | 2514 | 3047 | 177 | 2782 | 11 | 2 | 75 |
| 2002 | 775 | 2704 | 3479 | 299 | 2988 | 50 | 0 | 142 |
| 2003 | 687 | 2227 | 2914 | 1 | 2855 | 58 | 0 | 0 |
| 2004 | 711 | 2223 | 2934 | 1 | 2880 | 33 | 20 | 0 |
| 2005 | 925 | 3008 | 3933 | 0 | 3875 | 56 | 2 | 0 |
| 2006 | 1210 | 2455 | 3665 | 7 | 3624 | 34 | 0 | 0 |
| 2007 | 1158 | 3063 | 4221 | 1 | 4209 | 11 | 0 | 0 |
| 2008 | 987 | 2511 | 3498 | 3 | 3458 | 13 | 24 | 0 |
| 2009 | 811 | 2692 | 3503 | 0 | 3473 | 26 | 4 | 0 |
| 2010 | 1101 | 3046 | 4147 | 0 | 4124 | 23 | 0 | 0 |
| 2011 | 1157 | 3114 | 4271 | 0 | 4227 | 44 | 0 | 0 |
| 2012 | 1946 | 3573 | 5519 | 0 | 5366 | 126 | 27 | 0 |
| 2013 | 1505 | 3296 | 4801 | 0 | 4404 | 395 | 2 | 0 |
| 2014 | 1574 | 3829 | 5403 | 0 | 4281 | 1122 | 0 | 0 |
| 2015 | 1749 | 3449 | 5198 | 0 | 3836 | 1362 | 0 | 0 |
| 2016 | 1874 | 3011 | 4885 | 0 | 2935 | 1939 | 11 | 0 |
| 2017 | 1989 | 3323 | 5312 | 0 | 3638 | 1674 | 0 | 0 |
| Total | 21212 | 52518 | 73730 | 1238 | 64994 | 6982 | 106 | 410 |

The number of final-round matches analyzed for the 311 Go players was 21,212, as described in the Materials and Methods section. [a] The points added to the score of a player who played with white stones as a compensation for playing second in the game (*12, 13*).

[b] Final rounds comprised final matches in championships or special tournaments. [c] Zero points were given in master level-up matches only.

**Table S2. Magnetic field parameters for modulated GMF experiments.**

| MF | Player seat [a] | Components (µT) | | | Total Intensity (µT) | Relative total Intensity (%) [b] |
|---|---|---|---|---|---|---|
| | | X | Y | Z | | |
| Con | N | 30.7 | − 6.6 | 32.4 | 45.1 | 100 |
| | S | 29.3 | − 4.1 | 33.2 | 44.5 | 99 |
| ≈ 0 | N | 0.0 | 0.0 | 0.0 | 0.0 | 0 [c] |
| | S | 17.8 | − 5.8 | 34.4 | 39.2 | 88 [d] |
| Horizontal only mN | N | 44.7 | − 5.6 | 0.0 | 45.0 | 100 |
| | S | 37.6 | − 6.7 | 33.1 | 50.5 | 112 |
| Vertical inverted | N | 30.6 | − 6.5 | − 32.5 | 45.1 | 100 |
| | S | 32.4 | − 5.8 | 35.2 | 48.2 | 107 |
| mE | N | 0.0 | 45.1 | 0.0 | 45.1 | 100 |
| | S | 17.7 | − 8.2 | 34.5 | 39.6 | 88 |
| mW | N | − 0.1 | − 45.1 | 0.1 | 45.1 | 100 |
| | S | 17.0 | − 3.4 | 34.6 | 38.7 | 86 |
| mS | N | − 45.0 | 0.0 | 0.1 | 45.0 | 100 |
| | S | − 2.0 | − 6.1 | 33.6 | 34.2 | 76 |
| mNE | N | 42.0 | 16.2 | 0.2 | 45.0 | 100 |
| | S | 35.4 | − 6.2 | 35.2 | 50.3 | 112 |
| mNW | N | 30.3 | − 33.2 | 0.1 | 45.0 | 100 |
| | S | 32.1 | − 4.0 | 35.7 | 48.2 | 107 |
| Vertical only (down) | N | − 0.4 | 0.0 | 45.0 | 45.0 | 100 |
| | S | 15.2 | − 5.5 | 34.1 | 37.7 | 84 |
| Vertical only (up) | N | 0.2 | 0.1 | − 45.0 | 45.0 | 100 |
| | S | 15.7 | − 5.6 | 49.8 | 52.5 | 117 |

The intensities of the magnetic fields were the means from measurements on the glabella of the three subjects with the highest, median, and the lowest sitting height on the chair. [a] The north seat (N) and south seat (S) are located at the center and outside of the Helmholtz coils, respectively, on the geomagnetic north-south axis. [b] The relative total intensity of the S-seat magnetic field was compared to that of the corresponding N-seat in each GMF condition. [c,d] Relative total intensity at the N and S seats in the ≈ 0 condition was compared with the total intensity at the N and S seats in the Con condition, respectively. MF, magnetic field. MF conditions were named with respect to the N-seat; Con, the ambient GMF; ≈ 0, near-zero GMF; Horizontal only mN, the modulated magnetic field with horizontal component alone that directed toward the ambient magnetic north (mN); Vertical inverted, the ambient GMF with the inverted vertical component; mE, mW, and mS, the modulated magnetic field with horizontal component alone that directed toward the ambient magnetic east, west, and south, respectively; mNE and mNW, the modulated magnetic field with horizontal component alone that directed toward 30° magnetic northeast and 30° magnetic northwest, respectively, with respect to the ambient magnetic north; Vertical only (up) or Vertical only (down), the modulated magnetic field with vertical component alone that directed up or down, perpendicular to the surface of Earth.

**Other supporting files** (https://github.com/kschae5/arXiv)

**Movie S1 :** The stone choice process in Go games

**Movie S2 :** The stone choice procedure in the laboratory experiments

**Data S1 :** The data of Fig. S1

**Data S2 :** The data of Fig. 1

**Data S3 :** The data of Fig. 2

**Data S4 :** The data of Fig. 3

**Data S5 :** The data of Fig. 4

**Data S6 :** The data of Fig. S3

**Data S7 :** The data of Fig. S4

**Data S8 :** The data of Fig. 5

**Data S9 :** The record form and an example of the stone choice

**Data S10 :** The results of the Anderson–Darling test

**Data S11 :** The results of the two-sample *t*-test

**Data S12 :** The results of the percentile bootstrap analysis

**Data S13 :** The raw data for the percentile bootstrap analysis